\documentclass[a4paper,12pt]{article}

\begin{document}
\makeatletter
\renewcommand{\theequation}{\thesection.\arabic{equation}}
\@addtoreset{equation}{section}
\makeatother

\title{ Where Does Black Hole Entropy Lie? \\
-{\Large Some Remarks on Area-Entropy Law, Holographic Principle and Noncommutative Space-Time}}

\author{\large Sho Tanaka\footnote{Em. Professor of Kyoto University, E-mail: st-desc@kyoto.zaq.ne.jp }
\\[8 pt]
 Kurodani 33-4, Sakyo-ku, Kyoto 606-8331, Japan}
\date{}
\maketitle

\vspace{-10cm}
\rightline{}
\vspace{10cm}

\abstract{ In confrontation with serious and fundamental problems towards ultimate theory of quantum gravity and Planck scale physics, we emphasize the importance of underlying noncommutative space-time such as Snyder's or Yang's   Lorentz-covariant quantized space-time.  The background of Bekenstein-Hawking's Area-entropy law and  Holographic principle is now substantially understood in terms of  {\it Kinematical} Holographic Relation [KHR], which holds in Yang's quantized space-time as the result of the kinematical reduction of spatial degrees of freedom caused by its own 
nature of noncommutative geometry. [KHR] implies a proportional relation, $ n_{\rm dof} (V_d^L)= {\cal A} (V_d^L) / G_d$,  between the number of  spatial degrees of freedom  $n_{\rm dof} (V_d^L)$ inside any $d-$dimensional spherical volume $V_d^L$ with radius $L $ and its boundary area ${\cal A} (V_d^L).$  It yields a substantial basis for our new area-entropy law of black holes and further enables us to connect ``The First Law of  Black Hole Mechanics" with ``The Thermodynamics of Black Holes,"towards our final  goal: {\it Statistical} and {\it substantial} understanding of area-entropy law of black holes under a novel concept of noncommutative  quantized space-time.

\newpage
\section{\normalsize Introduction}

 In confrontation with serious and fundamental  problems towards {\it ultimate} theory of quantum gravity and Planck scale physics,  we emphasize the importance of noncommutative space-time such as Snyder's or Yang's Lorentz-covariant quantized space-time [1,2], [3,4]. As will be seen in what follows, the background of Bekenstein-Hawking's Area-entropy law or the so-called  Holographic principle, both underlying the present-day hot  issues around  black hole entropy, is now substantially understood directly or indirectly in terms of  {\it Kinematical} Holographic Relation [KHR]. The latter [KHR] was found [5,6] to hold in Yang's quantized space-time as the result of the definite kinematical reduction of spatial degrees of freedom caused by its own  nature of noncommutative geometry. (see section 3)
 
 On the other hand, it should be noted that, as was once pointed out by R.~Jackiw [7], the idea of noncommutative geometry or quantized space-time was first suggested by Heisenberg in the late 1930's so as to regulate the short-distance singularities in local quantum-field theories by virtue of the noncommutative-coordinate uncertainty.  In 1947, H. S. Snyder [1,2] accomplished  it successfully over a lot of challenges by prominent theoretical physicists in those days. Immediately after the Snyder's pioneering work, C.N.~Yang  proposed Yang's quantized space-time  [3,4],  in order to improve the Snyder's theory so as to satisfy the translation invariance in addition to the Lorentz-invariance. As will be reviewed in the next section, in particular, Yang's quantized space-time algebra [YSTA] is intrinsically equipped with short- and long-scale parameters, $\lambda$ (identifiable with Planck scale, see section 4.2) and $R,$\footnote{$R$ may be promisingly related to a {\it fundamental cosmological constant} in connection with the recent dark energy problem, as was preliminarily inferred in [9].} gives a finite number of spatial degrees of freedom for any finite spatial region without destroying the Lorentz-covariance and gives a possibility of field theory free from ultraviolet- and infrared-divergences [8],[9].  

 Now, let us  first briefly review the idea of [KHR]. It is expressed by
\begin{eqnarray} 
[KHR] \hspace{1cm}   n_{\rm dof} (V_d^L)= {\cal A} (V_d^L) / G_d,
\end{eqnarray}
that is, the proportional relation between $n_{\rm dof}(V_d^L)$ and ${\cal A} (V_d^L)$ with proportional constant $G_d$ (see (3.5)-(3.6) in section 3), where $n_{\rm dof} (V_d^L)$ and ${\cal A}(V_d^L)$, respectively, denote the number of degrees of freedom of any $d$ dimensional bounded spatial region  $V_d^L$ with radius $L$ in Yang's quantized space-time, and the boundary area of its region in unit of Planck length.

 As a matter of fact, the possibility of the kinematical reduction of spatial degrees of freedom in noncommutative space-time may be well understood intuitively, 
in terms of the familiar {\it quantum correlation} among different components of spatial coordinates constrained to satisfy the noncommutative relations. Indeed, it yields a possibility of giving a simple clue for resolving the long-pending problem encountered in the Bekenstein-Hawking area-entropy relation [10]-[16] and the holographic principle [17]-[25], that is, the apparent gap between the degrees of freedom of any bounded spatial region associated with entropy and of its boundary area.

 As is easily imagined, the kinematical holographic relation [KHR] shown in Eq. (1.1) suggests that the entropy of any statistical system realized in the spatial region  $V_d^L$  must be proportional not to $V_d^L$, but  $ {\cal A} (V_d^L) $, namely, it yields  a new  area-entropy law. In other words, [KHR] strongly suggests our final goal of the present paper: {\it Statistical} and {\it substantial} understanding of area-entropy law of black holes under a novel concept of noncommutative space-time or quantized space-time.  

 Indeed, on the basis of this [KHR] (1.1) mentioned above, we derive in section 4 the following form of a new {\it area-entropy relation}  of black hole in a purely {\it statistical} way, through a simple  $D_0$ brane gas model [26]-[31] in Yang's quantized space-time  (see  (4.20) )
\begin{eqnarray}
 S_S(V_3^{R_S}) = {\cal A}\ (V_3^{R_S})\ {S_S[site] \over 4 \pi}.
\end{eqnarray}
The relation, further compared with the familiar Bekenstein proposal $ S_{BH} = \eta \ A  $, gives us a following  important relation with respect to Bekenstein parameter  $\eta$  

\begin{eqnarray}
\eta = {S_S[site] \over 4 \pi},
\end{eqnarray}
leaving its detailed explanation to  section 4 (see (4.21) ). The physical implication of the relation (1.3) or $S_S[site]$ will be argued in the final section in terms of {\it universality}  of black holes.

 The present paper is organized as follows.  In order to make the present paper as self-contained as possible, let us first simply review in sections 2-4 our present approach: In section 2 and section 3, respectively,  we briefly recapitulate Yang's quantized space-time and the derivation of the  kinematical holographic relation [KHR] mentioned above. In section 4, we examine in a compact way the area-entropy problem arriving at Eqs. (1.2) and (1.3) through a simple $D_0$ brane gas model developed in  Yang's quantized space-time. In section 5, we discuss the connection between ``The first law of  black hole mechanics" and ``The thermodynamics of black holes" through [KHR] and our new area-entropy relation mentioned above . The final section is devoted to Concluding Remarks and Further Outlook.

\section{\normalsize Yang's Quantized Space-Time Algebra [YSTA]  and Its Representations}  
  
We shall first review the Lorentz-covariant Yang's quantized space-time [3, 4] and its representation.  First, $D$-dimensional Yang's quantized space-time algebra [YSTA] was introduced  as the 
result of the so-called Inonu-Wigner's contraction procedure with two contraction parameters,  long $R$ and short $\lambda$, from $SO(D+1,1)$ algebra with generators $\hat{\Sigma}_{MN}$ [9] ; 
\begin{eqnarray} 
 \hat{\Sigma}_{MN}  \equiv i (q_M \partial /{\partial{q_N}}-q_N\partial/{\partial{q_M}}),
\end{eqnarray}
which work on $(D+2)$-dimensional parameter space  $q_M$ ($M= \mu,a,b)$ satisfying  
\begin{eqnarray}
             - q_0^2 + q_1^2 + \cdots + q_{D-1}^2 + q_a^2 + q_b^2 = R^2.
\end{eqnarray}
Here, $q_0 =-i q_D$ and $M = a, b$ denote two extra dimensions with space-like metric signature.

$D$-dimensional space-time and momentum operators, $\hat{X}_\mu$ and $\hat{P}_\mu$, 
with $\mu =1,2,\cdots,D,$ are defined by
\begin{eqnarray}
&&\hat{X}_\mu \equiv \lambda\ \hat{\Sigma}_{\mu a}
\\
&&\hat{P}_\mu \equiv \hbar /R \ \hat{\Sigma}_{\mu b},   
\end{eqnarray}
together with $D$-dimensional angular momentum operator $\hat{M}_{\mu \nu}$
\begin{eqnarray}
   \hat{M}_{\mu \nu} \equiv \hbar \hat{\Sigma}_{\mu \nu}
\end{eqnarray} 
and the so-called reciprocity operator
\begin{eqnarray}
    \hat{N}\equiv \lambda /R\ \hat{\Sigma}_{ab}.
\end{eqnarray}
Operators  $( \hat{X}_\mu, \hat{P}_\mu, \hat{M}_{\mu \nu}, \hat{N} )$ defined above 
satisfy the so-called contracted algebra of the original $SO(D+1,1)$, or Yang's 
space-time algebra:
\begin{eqnarray}
&&[ \hat{X}_\mu, \hat{X}_\nu ] = - i \lambda^2/\hbar \  \hat{M}_{\mu \nu}
\\
&&[\hat{P}_\mu,\hat{P}_\nu ] = - i\hbar / R^2\  \hat{M}_{\mu \nu}
\\
&&[\hat{X}_\mu, \hat{P}_\nu ] = - i \hbar \hat{N} \delta_{\mu \nu}
\\
&&[ \hat{N}, \hat{X}_\mu ] = - i \lambda^2 /\hbar \  \hat{P}_\mu
\\
&&[ \hat{N}, \hat{P}_\mu ] =  i \hbar/ R^2\ \hat{X}_\mu,
\end{eqnarray}
with other familiar relations concerning  ${\hat M}_{\mu \nu}$'s omitted.

Next, let us review briefly the representation [9] of YSTA. First of all, it is important to notice the following elementary fact that ${\hat\Sigma}_{MN}$ 
defined in Eq. (2.1) with $M, N$ being the same metric signature have discrete eigenvalues, i.e., $0,\pm 1 ,
 \pm 2,\cdots$, and those with $M, N$ being opposite metric signature have continuous eigenvalues. Therefore, from (2.3) and (2.4), one sees that ${\hat X}_u $ and ${\hat P}_u$ $(u = 1, 2, \cdots, D-1)$ have discrete eigenvalues in units of $ \lambda $ and ${\hbar}/R$, respectively, and both ${\hat X}_0 $ and ${\hat P}_0 $ continuous eigenvalues, consistently with the covariant commutation relations of YSTA.  This fact was first emphasized by Yang [3,4] with respect to the Snyder's quantized space-time mentioned above. 

As a matter of fact, this kind of eigenvalue structure of YSTA, in particular, the discrete          eigenvalue structure mentioned above is clearly indispensable for realizing  quantum field theories free from infrared- and ultraviolet-divergences on the one hand, but it is narrowly permitted consistently with Lorentz-covariance because of the fact that 
YSTA is  subject to Lorentz covariant {\it noncommutative} geometry on the other hand.  This conspicuous aspect is well understood by means of the familiar example of the three-dimensional angular momentum in quantum mechanics, where individual components are able to have discrete eigenvalues, consistently with the three-dimensional rotation-invariance, because of the fact that the individual components are {\it noncommutative} among themselves. 
 
Yang's space-time algebra [YSTA] presupposes for its representation space 
to take representation bases like 
\begin{eqnarray}
\hspace{-0.5cm} | t/\lambda, \ n_{12}, \cdots \rangle \equiv |{\hat{\Sigma}}_{0a} =t/\lambda \rangle \  |{\hat{\Sigma}}_{12}=n_{12}\rangle \cdots |{\hat{\Sigma}}_{910}=n_{910}\rangle,
\end{eqnarray}
where $t$ denotes {\it time}, the continuous eigenvalue of $\hat{X}_0 \equiv \lambda\ \hat{\Sigma}_{0 a}$ 
and $n_{12}, \cdots$ discrete eigenvalues of maximal commuting set of subalgebra of $SO(D+1,1)$ which are 
commutative with ${\hat{\Sigma}}_{0a}$, for instance, ${\hat{\Sigma}}_{12}$, ${\hat{\Sigma}}_{34},\cdots , 
{\hat{\Sigma}}_{910}$, when $D=11.$[9],[32]

Indeed, an infinite dimensional linear space expanded by $|\ t/\lambda, 
n_{12},\cdots \rangle$ mentioned above provides a representation space of unitary infinite dimensional representation of YSTA. 
It is the so-called ``quasi-regular representation"[33] of SO(D+1,1),\footnote{It corresponds, in the case of 
unitary representation of Lorentz group $SO(3,1)$, to taking $K_3\ (\sim \Sigma_{03})$ and $J_3\ (\sim \Sigma_{12})$ 
to be diagonal, which have continuous and discrete eigenvalues, respectively, instead of ${\bf J}^2$ and $J_3$ in 
the familiar representation.}
and is decomposed into the infinite series of the ordinary unitary irreducible representations of 
$SO(D+1,1)$ constructed on its maximal compact subalgebra, $SO(D+1)$.
 (See  Chapter {\bf 10 },  10.1. ``Decompositions of Quasi-Regular Representations and Integral Transforms" in ref. [33] ) 

It means that there holds the following form of decomposition theorem,
\begin{eqnarray}
\hspace{1cm} | t/\lambda, n_{12},\cdots \rangle = \sum_{\sigma 's}\ \sum_{l,m}\  C^{\sigma's, n_{12}, \cdots }_{l,m}(t/\lambda)\ 
 | \sigma 's ; l,m \rangle,
\end{eqnarray}      
with expansion coefficients $C^{\sigma's, n_{12}, \cdots}_{l,m}(t/\lambda).$  In Eq.(2.13), 
$|\sigma 's ; l, m \rangle's $ on the right hand side describe the familiar unitary irreducible representation 
bases of $SO(D+1,1)$, which are designated by $\sigma 's$ and $(l,m),$  
\footnote{In the familiar unitary irreducible representation of $SO(3,1)$, it is well known that $\sigma$'s are 
represented by two parameters, $(j_0, \kappa)$, with $j_0$ being $1,2, \cdots \infty$ and $\kappa$ being purely 
imaginary number, for the so-called principal series of representation. With respect to the associated representation 
of $SO(3)$, when it is realized on $S^2$, as in the present case, $l$'s denote positive integers, 
$l= j_0, j_0+1, j_0+2,\cdots,\infty$, and $m$ ranges over $\pm l, \pm(l-1) , \cdots,\pm1, 0.$ } 
denoting, respectively, the irreducible unitary representations of $SO(D+1,1)$ and the associated 
irreducible representation bases of $SO(D+1)$, the maximal compact subalgebra of $SO(D+1,1)$, mentioned above. 
It should be noted here that, as remarked in [9], $l$'s  are limited to be integer, excluding the possibility of 
half-integer, because of the fact that generators of $SO(D+1)$ in YSTA are defined as differential operators 
on $S^D$, i.e., ${q_1}^2 + {q_2}^2 + \cdots + {q_{D-1}}^2 + {q_a}^2 + {q_b}^2 = 1.$

In what follows, let us call the infinite dimensional representation space introduced above for the representation of 
YSTA, Hilbert space I, in distinction to Hilbert space II, that is, Fock-space constructed dynamically by 
creation-annihilation operators of second-quantized fields in  Yang's quantized space-time (see section 4).

\section{\normalsize Derivation of  Kinematical Holographic Relation [KHR]}

Let us show briefly the derivation process of our central concern,  [KHR]  in Yang's quantized space-time mentioned in Introduction, that is,  $ [KHR]\ n_{\rm dof} (V_d^L)= {\cal A} (V_d^L) / G_d \  (1.1), $ leaving its detail to [5].

The boundary surface of any $d-$dimensional spatial sub-region  $V_d^L$ with radius $L$, in $D-$ dimensional Yang's quantized space-time is defined by 

\begin{eqnarray}  
{{\hat X}_1}^2 + {{\hat X}_2}^2 + \cdots + {{\hat X}_d}^2 = L^{2}.
\end{eqnarray}

The boundary area of $V_d^L$ in the unit of $\lambda$, that is,  ${\cal A}\ (V_d^L)$  defined by $``{\rm Area\ of}\ S^{d-1} {\rm with\ radius}\ L/\lambda" $ is given as follows 
  
\begin{eqnarray}
\hspace{-1cm}{\cal A}\ (V_d^L) ={(2 \pi)^{d/2} \over {(d-2)!!}} (L/\lambda)^{d-1} 
&& {\rm for\ {\it d}\ even}
\nonumber\\                   =2 {(2\pi)^{(d-1)/2} \over {(d-2)!!}} (L/\lambda)^{d-1} &&{\rm for\ {\it d}\ odd}. 
\end{eqnarray}

On the other hand, the number of degrees of freedom of $V_d^L$,  $n_{\rm dof} (V_d^L)$,   is found in a certain irreducible 
representation of $SO(d+1)$, a minimum subalgebra of Yang quantized space-time which includes the $d$ spatial coordinate operators, $\hat{X}_1, \hat{X}_2, \cdots, \hat{X}_d$ in Eq. (3.1) needed to properly describe $V_d^L$. The  $SO(d+1)$ is really constructed by the generators $\hat{\Sigma}_{MN}$ with $M,N$ ranging over $a,1,2, \cdots d.$ Let us denote the irreducible representation by $\rho_l\ (V_d^L)$ with the characteristic integer $l$ which indicates the maximal eigenvalue of any generators,  $\hat{\Sigma}_{MN},$ of $SO(d+1).$

It is well known that the irreducible representation of $SO(d+1)$,  $\rho_l(V_d^L)$ with the characteristic integer $l$ mentioned above is simply given in the algebraic method
[34], on the representation space $S^d =SO(d+1)/SO(d)$, by taking the subalgebra $SO(d)$, $\hat{\Sigma}_{MN}$ with $M,N$ ranging over $1,2,\cdots,d,$ for instance. Furthermore, one finds that the representation $\rho_{[L/\lambda]} (V_d^L)$ by taking $ l = [L/\lambda],$  i.e., the nearest integer to $ L/\lambda $, just {\it properly} describes all of generators of $SO(d+1)$ inside the bounded spatial region $V_d^L$, because  $[L/\lambda ]$ is the largest eigenvalue of any generators of $SO(d+1)$ in the representation $\rho_{[L/\lambda]}$.
 As a result, one finds that the {\it dimension} of $\rho_{[L/\lambda]}$ can be reasonably identified with the spatial degrees of freedom inside $V_d^L$, that is, 
$ n_{\rm dof} (V_d^L) = {\rm dim} ( \rho_{[L/\lambda]} (V_d^L)).$

According to the Weyl dimension formula applied to the irreducible representation of  $SO(d+1)$, the dimension of $\rho_l$ is given by [34]

\begin{eqnarray}
{\rm dim}(\rho_l) = {l +\nu \over \nu}\  {(l+2\nu -1) ! \over l !( 2\nu - 1)!},
\end{eqnarray}
with $ \nu = (d-1)/2$ in the case $ d\geq 2$ (see, more in detail, [5]).   

One immediately finds that 
\begin{eqnarray}
 n_ {\rm dof}\ (V_d^L)  &&= {\rm dim}\ ( \rho_{[L/\lambda]}\ (V_d^L)) \nonumber \\
&& = {2[L/\lambda]+d-1 \over [L/\lambda]} {([L/\lambda] +d-2)! \over ([L/\lambda]-1)! (d-1)!} \nonumber \\     
&&\sim { 2 \over (d-1) !} [L/\lambda]^{d-1} ,
\end{eqnarray}
where the expression in the last line holds in the case $ [L/\lambda] \gg d.$
Thus, comparing Eqs. (3.2) and (3.4), we finally arrive at the following kinematical holographic relation;

\begin{eqnarray}
[KHR] \hspace{2cm}       n_{\rm dof}  (V_d^L)=
{\cal A} 
 (V_d^L) / G_d
\end{eqnarray}
with the proportional constant $G_d;$
\begin{eqnarray}
G_d  &&\sim\   {(2 \pi)^{d/2} \over 2} (d-1)!! \qquad {\rm for\ {\it d}\  even} \nonumber\\
      &&\sim\  (2 \pi)^{(d-1)/2} (d-1)! !\quad    {\rm for\   {\it d}\  odd}.
\end{eqnarray}

\section{\normalsize Statistical  Derivation of Area-Entropy Relation based on [KHR] in Yang's Quantized Space-time}

 In this section, we show that the kinematical holographic relation [KHR]  (3.5) derived in the previous section successfully leads us  to the {\it statistical} derivation of a new  area-entropy relation through a simple $D_0$ brane gas model formed inside $V_d^L$  in Yang's quantized space-time. The main cause of succes is easily understood from its own form of  [KHR], so far as the entropy of the statistical system is to be proportional to  $ n_{\rm dof}  (V_d^L),$ as seen below.

\subsection{\normalsize Statistical Derivation of Area-Entropy Relation [AER] and  Area-Mass Relation [AMR]  in $D_0$ brane Gas System}

The  $D_0$ brane gas system formed inside $V_d^L$ is most likely described in terms of the {\it second-quantized field} of $D_0$ brane [6, 32, 36] or the $D$ particle [26] defined in Yang' quantized space-time,  $V_d^L$.   First of all, according to the argument given in section 3, the spatial structure of $V_d^L$ is described through the specific representation $ \rho_{[L/\lambda]}\ (V_d^L)$. Let us denote its orthogonal 
basis-vector system in the representation space, which we called Hilbert space I at the end in section 2  in distinction to Hilbert space II, as follows 
\begin{eqnarray} \rho_{[L/\lambda]}\ (V_d^L): \quad |\ m\rangle,  \qquad  m= 1,2,\cdots, n_{\rm dof}(V_d^L).     
\end{eqnarray}   

Meanwhile, one should notice that the {\it second quantized} $D_0 $ brane  field mentioned above, that is,  ${\hat {\bf D}}_0$, must be the linear operators operating on Hilbert space I, and described by $n_{\rm dof}(V_d^L) \times n_{\rm dof}(V_d^L)$ 
matrix under the representation $\rho_{[L/\lambda]} (V_d^L)$ like $\langle m\ |{\hat {\bf D}}_0|\ n \rangle$ on the one hand.  On the other hand, each matrix element must be operators operating on Hilbert space II, playing the role of creation-annihilation of $D_0$ branes. On the analogy of the ordinary quantized local field, let us define those creation-annihilation operators through the diagonal parts in the following way:\footnote{On the other hand, 
the non-diagonal parts, $\langle m\ |{\hat {\bf D}}_0|\ n \rangle,$ 
are to be described in terms like ${\bf a}_m {\bf a}_n^\dagger$ or ${\bf a}_m^\dagger {\bf a}_n$ in accord with the idea of M-theory where they are conjectured to be concerned with the interactions between $[site\ m]$ and $[site\ n].$  The details must be left to the rigorous study of the second quantization of $D_0$-brane field. [36] }  
\begin{eqnarray}
\langle m\ | {\hat{\bf  D}_0}|\ m \rangle\  \sim\ {\bf a}_m\ {\rm or}\  {\bf a}_m^\dagger.
\end{eqnarray}
In the above expression, ${\bf a}_m$ and ${\bf a}_m^\dagger $ denote annihilation and creation 
operators of $D_0$ brane, respectively, and satisfy the familiar commutation relations,
\begin{eqnarray}
&&[{\bf a}_m, {\bf a}_n^\dagger]= \delta_{mn},
\\
&&[{\bf a}_m, {\bf a}_n]= 0 .
\end{eqnarray}
The labeling number $m$ of basis vectors, which ranges from $1$ to $n_{\rm dof}(V_d^L),$ 
plays the role of {\it spatial coordinates} of $V_d^L$ in the present noncommutative YSTA, corresponding to the lattice point in the lattice theory. Let us denote the {\it point} hereafter $[site]$ or $[site\ m]$ of $V_d^L$.

We focus our attention on quantum states constructed dynamically in Hilbert space II by the creation-annihilation 
operators ${\bf a}_m$ and ${\bf a}_m^\dagger$ of ${\bf {\hat D}}_0$ brane field introduced above at each [site] inside $V_d^L$. One should 
notice here the important fact that in the present simple $D_0$ brane gas model which ignores all interactions of $D_0$ branes, 
each $[site]$ can be regarded as independent quantum system and can be described in general by own statistical operator, while 
the total system of gas is described by their direct product. In fact, the statistical operator at each $[site\  m]$ 
denoted by ${\bf W}[m]$, is given in the following form,
\begin{eqnarray} 
{\bf W}[m] = \sum_ k  w_k\  |\ [m]: k \rangle\  \langle k :[m]\ |,
\end{eqnarray}
with 
\begin{eqnarray}
 |\ [m]:  k \rangle  \equiv {1 \over \sqrt{k!}}({\bf a}_m^\dagger)^{k}|\ [m]:0\rangle.
\end{eqnarray}
That is, $|\ [m]:  k \rangle \ ( k=0, 1, \cdots)$ describes the normalized quantum-mechanical state in Hilbert space II with 
$k$ $D_0$ branes constructed by ${\bf a}_m^\dagger$ on $|\ [m]:0\rangle,$ i.e., the vacuum state of $[site\ m]$.\footnote{The proper vacuum state in Hilbert space II is to be expressed by their direct product.} And $w_k$'s denote 
the realization probability of state with occupation number $k$, satisfying $\sum_k w_k = 1.$

We assume that the  $D_0$ brane gas system is under an equilibrium state with the equilibrium temperature $T$ and the statistical operator at each $[site\ m]$ is common to every [site] with the common values of $w_k$'s :
\begin{eqnarray}
 w_k = e^{-\mu k/T} / Z(T),
\end{eqnarray}
where
\begin{eqnarray}
Z(T) \equiv \sum_{k=0}^\infty e^{-\mu k/T} = 1 / (1- e^{-\mu / T}).
\end{eqnarray}
In the above expression,  $\mu$ denotes the average energy or effective mass of the individual $D_0$ brane inside $V_d^L$. 

 The statistical operator of total system in $V_d^L$, ${\bf W}(V_d^L)$, is now given by  
\begin{eqnarray}
{\bf W}(V_d^L) = {\bf W}[1] \otimes {\bf W}[2] \cdots \otimes {\bf W}[m] \cdots \otimes {\bf W}[n_{dof}].
\end{eqnarray}

Consequently, one finds that the entropy and the energy or effective mass  of the total system,  $S(V_d^L)$ and $M(V_d^L)$ are respectively given by 
\begin{eqnarray}
 S(V_d^L) = - {\rm Tr}\ [{\bf W}(V_d^L)\ {\rm ln} {\bf W}(V_d^L)] = n_{\rm dof}(V_d^L)\times S[site], 
\end{eqnarray}
and
\begin{eqnarray}   
M(V_d^L) = \mu {\bar N}[site]\  n_{\rm dof}(V_d^L)\ (\sim  \mu {\bar N}[site] {2  \over (d-1)!}  [L/\lambda]^{d-1}).
\end{eqnarray}
In the above expressions, $S[site]$ in Eq. (4.10) denotes the entropy of each [site] assumed to be common to every [site] and is given by
\begin{eqnarray}
S[site] &&\equiv - \sum_k w_k\  {\rm \ln} w_k = {\mu {\bar N}[site] \over T}  + \ln  Z(T)
\nonumber \\
 &&= - \ln (1 - e^{- \mu /T}) + {\mu \over T}\ ( e^{\mu /T}- 1)^{-1},
\end{eqnarray}
and  ${\bar N}[site]$ in Eqs. (4.11) and (4.12  ) denotes the average occupation number of $D_0$ brane at each $[site]$
\begin{eqnarray}
{\bar N}[site] \equiv \sum_k k w_k = (\ e^{\mu / T}- 1)^{-1}.
\end{eqnarray}

Comparing Eq. (4.10) with [KHR]\ (3.5) derived in the preceding section, we find an important fact that the 
entropy $S(V_d^L)$ is proportional to the surface area ${\cal A}\ (V_d^L)$, that is, a new area-entropy relation [AER] is derived in a purely {\it statistical} way through [KHR],
\begin{eqnarray}
\hspace{-3cm} [AER] \hspace{2cm}         S(V_d^L) = {\cal A}\ (V_d^L)\ {S[site] \over G_d},
\end{eqnarray}
where $G_d$ is given by (3.6). 

Furthermore, comparing (4.11) with [KHR]  (3.5) and (4.10), respectively, we obtain  a area-mass relation ([AMR])
\begin{eqnarray}
\hspace{-3cm} [AMR] \hspace{2cm} M(V_d^L) = {\cal A}(V_d^L)\ {\mu {\bar N}[site] \over G_d}
\end{eqnarray}
and a mass-entropy relation ([MER]) 
\begin{eqnarray}
\hspace{-2cm} [MER] \hspace{2cm}         M(V_d^L) / S(V_d^L) = \mu {\bar N}[site] / S[site].
\end{eqnarray}

At the end of this section, let us notice that there holds the following relation between 
$S [site]$ and ${\bar N}[site]$,   
\begin{eqnarray}
S [site] = \ln (1+ {\bar N}[site]) + {\bar N} [site] \ln (1+ {\bar N}^{-1}[site]),
\end{eqnarray}
which comes from Eqs. (4.12) and (4.13). The relation will be used in the final section  in terms of {\it universality} of black holes.

\subsection{\normalsize Schwarzschild Black Hole and Area-Entropy Relation in $D_0$ brane Gas System}

Let us consider how the present gas system transforms into a black hole. We assume here that the system is under  $ D=4\  ( d=3)$, and transforms into a Schwarzschild black hole, in which the quantities $ \mu$, ${\bar N}[site]$ and $S  [site]$ introduced in the previous sub-section 4.1 acquire certain limiting 
values, $ \mu_S$, ${\bar N}_S[site]$ and $S_S [site]$, while the size of the system, $L$, becomes $R_S$, that is, the Schwarzschild radius given by 
\begin{eqnarray}   
R_S =  2 G M_S(V_3^{R_S})/c^2.
\end{eqnarray}
In the above expression, $G$ and $c$ denote Newton's constant and the velocity of light, respectively, and $M_S(V_3^{R_S})$ is given by Eq. (4.11) with $L= R_S$, $\mu = \mu_S$ and ${\bar N}[site] = {\bar N}_S[site]$.

 Indeed, inserting the above values into the last expression of Eq.(4.11), we arrive at the important relation, called hereafter the black hole condition [BHC], 
\begin{eqnarray}
[BHC] \hspace{0.5cm} M_S(V_3^{R_S}) = {\lambda^2 \over 4\mu_S {\bar N}_S [site]}{c^4 \over G^2} 
= {M_P^2 \over 4 \mu_S {\bar N}_S[site]}.
\end{eqnarray}
Here, we assumed that $\lambda$, i.e., the short scale parameter in YSTA is equal  
to Planck length $l_P = [G \hbar / c^3]^{1/2} = \hbar /( c M_P )$, where $M_P$  denotes Planck mass. In what follows, we will use Planck units in $D=4 $, with $M_P = l_P = \hbar = c = k =1$.

Then, we simply obtain the area-entropy relation [AER] under the Schwarzschild black hole by 
inserting the above limiting values into [AER] (4.14)   
\begin{eqnarray} 
[AER] \hspace{1.5cm}  S_S(V_3^{R_S}) = {\cal A}\ (V_3^{R_S})\ {S_S[site] \over 4 \pi},
\end{eqnarray}
noticing that $G_{d = 3} =4 \pi.$

It is quite interesting to note that, by comparing the above relation  [AER] with the famous Bekenstein proposal $S_{BH}= \eta A$, where  $S_{BH} = S_S(V_3^{R_S}) $ and $A=  {\cal A}\ (V_3^{R_S})$ in our present scheme, the Bekenstein parameter $\eta$ is expressed as 
\begin{eqnarray}
\eta = {S_S[site] \over 4 \pi}.
\end{eqnarray}
The  physical implication of the relation will be argued in terms of {\it universality}  of black holes in the final section 6.

Hereupon, it is a very important problem how to relate the above area-entropy relation [AER] (4.20) under a Schwarzschild black hole state with [AER] (4.14) of $D_0$ brane gas system in general, which is derived irrelevantly of the detail whether the system is under a black hole state or not. The problem certainly exceeds the limits of applicability of the present simple model of $D_0$ brane gas, where the system is treated solely as a {\it static} state under {\it given} values of parameters, $\mu$, ${\bar N}[site]$ and $S[site]$, while the critical behavior around the formation of Schwarzschild 
black hole must be formed under a possible ${\it dynamical}$ change of their values. 

In order to supplement such a defect of the present static toy model, let us try here a Gedanken-experiment, 
in which one increases the entropy of the gas system $S(V_3^L)$, keeping its size $L$ at the initial value $L_0$, 
until the system changes into to a Schwarzschild black hole, where (4.20) with $R_S = L_0$ holds. Then, one 
finds that according to [AER] (4.14), the entropy of $[site]$, i.e.,  $S[site]$ (starting from any initial value $S_0[site]$) increases proportionally to $S(V_3^L)$  under the fixed $L=L_0$ and reaches the limiting value $S_S [site]$,  because ${\cal A}(V_3^L)$ with the fixed $L=L_0$ in (4.14) is invariant during the process. Namely, one finds a very simple fact 
that $S_0[site] \leq S_S [site].$  However, this simple fact combined with [AER] (4.14) leads us to the following  form of  {\it Holographic entropy bound} [HEB] or {\it Spherical entropy bound} [SEB] [25]   

\begin{eqnarray}
[HEB] \hspace{2cm}  S(V_3^L) \leq {{\cal A} (V_3^L) S_S[site] \over 4 \pi}
\end{eqnarray}
\begin{eqnarray}
[SEB] \hspace{2cm} S(V_3^L) \leq \eta {\cal A} (V_3^L),
\end{eqnarray}
where the equality holds for Schwarzschild black hole, as seen in Eq. (4.20).

Finally, we consider in our present scheme the Hawking radiation temperature $T_{H.R.}$ of the gas system under Schwarzschild black hole, which is defined by 
\begin{eqnarray}
T_{H.R.}^{-1} =  dS_S/dM_S.
\end{eqnarray}
Noticing the relation [AER] (4.20) with $ {\cal A} = 4 \pi  R_S^2 =16 \pi M_S^2$, we immediately 
find 
\begin{eqnarray} 
T_{H.R.}^{-1} ={d \over dM_S} S_S  = {d \over d M_S} (16 \pi M_S^2 S_S[site]/{4 \pi})     \cr
\nonumber\\             = 8 M_S S_S[site] + 4M_S^2  {d \over d M_S}{S_S[site]}.
\end{eqnarray}

Here, it is quite important to notice the relation (4.21), where $S_S[site]$ is related to  Bekenstein parameter $\eta$, i.e., $S_S[site] = 4\pi \eta$. That is,  it  tells us that $S_S[site]$ possesses some kind of  {\it universal} nature independent of individual black holes with different masses,  as will be discussed in the final section. Then, one immediately finds that the second term in the last expression 
of (4.25) becomes vanishing,  and arrives at 
\begin{eqnarray}
T_{H.R.} (= 1/(8 M_S S_S[site])) = {\kappa \over {8 \pi \eta}},    
\end{eqnarray}
where $\kappa ( \equiv  1/(4 M_S)) $ is the so-called surface gravity of black hole and  $T_{H.R.}$ tends to the familiar value  ${\kappa \over 2 \pi}$ if one takes $\eta =1/4$.

\section{\normalsize  ``Where Does Black Hole Entropy Lie?"- The first law of black hole mechanics and The thermodynamics of black holes }

In the previous section,  we have shown in detail  that the statistical system developed over  $V_d^L$ in the Yang's quantized space-time can be described
in Hilbert space I with  representation bases of $\rho_{[L/\lambda]}\ (V_d^L)$
  
\begin{eqnarray} \rho_{[L/\lambda]}\ (V_d^L): \quad |\ m \rangle,\quad   m= 1,2,\cdots, n_{\rm dof} (V_d^L).     
\end{eqnarray}    
So we  can now simply  say that black hole entropy lies, {\it statistically} and  {\it substantially}, inside $V_d^L$ with degrees of freedom $n_{\rm dof}(V_d^L)$ in  Yang's quantized space-time, satisfying the area-entropy law  without any contradiction, although the argument in section 4 was put forwarded  through a simple $D_0$ brane gas model on the basis of [KHR]. 
 
At this point,  let us consider here the central issues around  ``The First Law of Black Hole Mechanics" and  ``The Thermodynamics of Black Holes"  in connection with the title of the present section  ``Where Does Black Hole Entropy Lie?"

It is well known that  Bardeen, Carter and Hawking  early presented ``The integral and differential  mass formula," under the title ``The Four Laws of Black Hole Mechanics." (1973) [39] They derived the formula through a stationary axisymmetric solution of the Einstein equations containing black hole.  They presented the differential mass formula
of the following form:  

\begin{eqnarray}
\hspace{1cm} \delta M =\int \Omega \delta dJ + \int {\bar \mu} dN + \int {\bar \theta} \delta dS 
+\Omega_H \delta J_H + {\kappa \over 8\pi} \delta A.
\end{eqnarray} 
In the above expression,   $M$  describes the mass measured from infinity, $\Omega_H $ and $J_H$, respectively, angular velocity and angular momentum of black hole.    

It is also well-known that the last term  $ {\kappa \over 8\pi } \delta A $ of the differential mass formula (5.2), which comes from 2-surface boundary $\partial B $ at the event horizon of black hole  with surface gravity $\kappa$ and the horizon area,  $A$, gives rise to the origin of the present-day hot arguments around black holes.[40-42] 

Among these arguments,  Wald  arrived at the idea, ``Black Hole Entropy is Noether Charge" (1993) [43] with respect to the last term $ { \kappa \over 8\pi } \delta A $ mentioned above. We highly notice his subsequent works,  for instance,  ``Gravitation, Thermodynamics, and Quantum Theory"(1999) [44] or ``The Thermodynamics of Black Holes "(2001) [45] in which he endeavors consistently to connect {\it The First Law of  Black Hole Mechanics} with {\it The Thermodynamics of Black Holes.}   

One should notice, however, that ``The Thermodynamics of Black Holes"(2001) [45] mentioned above encounters a severe question ``What (and Where) are the Degrees of Freedom responsible for Black Hole Entropy" ( Open Issues in section 6), and further  ``Gravitation, Thermodynamics, and Quantum Theory"[44] also raises "Some unsolved issues and puzzles"( see section 4)  and is closed with the following impressive sentence; `` I believe that the above puzzle suggests that we presently lack the proper conceptual framework with which to think about entropy in the context of general relativity$\cdots.$"

Meanwhile,  we find out that ``The First Law of Black Hole Mechanics" is well matched with our present scheme,  on account of the fact that the last term $ { \kappa \over 8\pi } \delta A $ in Eq. (5.2) is now simply rewritten by using the kinematical holographic relation [KHR] (3.5) in the following way
\begin{eqnarray}
{ \kappa \over 8 \pi } \delta A  = {\kappa G_3 \over 8\pi } \delta n_{\rm dof} ( V_3^{R_S}) = {\kappa \over 2} \delta n_{\rm dof} ( V_3^{R_S}),
\end{eqnarray}
where  $G_3 = 4\pi $ and   $V_3^{R_S}$  corresponds to  3-dimensional hypersurface  inside the black hole. One sees that at this stage  ${ \kappa \over 8\pi } \delta A$ is  related through [KHR] with variation of the spatial degrees of freedom {\it inside} black hole  $\delta n_{\rm dof}( V_3^{R_S})$ beyond the boundary of black hole ${\partial B}$, but  still irrelevantly to ``Thermodynamics of Black Holes."       

Indeed,``The First Law of  Black Hole Mechanics" can be logically and actually connected with ``The Thermodynamics of Black Holes" when  and only when the last term  $  {\kappa \over 8\pi} \delta A  $ in (5.2) is transformed further in the following form 
\begin{eqnarray}
 {\kappa \over 8\pi } \delta A && = {\kappa \over 2}\  \delta\ ( {S_S (V_3^{R_S}) \over S_S [site]})
\nonumber \\
      && =  {\kappa \over{8\pi \eta}} \delta S_S(V_3^{R_S}) =T_{H.R.}\  \delta S_S (V_3^{R_S}),
\end{eqnarray}
by means of the relation (4.20).  In the last expression in (5.4),  Eqs. (4.21) and (4.26) are used, that is, $S_S[site]=4\pi \eta$ and $T_{H.R.} ={ \kappa \over 8\pi \eta}.$  
 
We close this section in answer to the question "Where does black hole entropy lie?" by emphasizing again that  black hole entropy {\it statistically} and  {\it substantially} lies inside $V_3^{R_S}$ in Yang's quantized space-time, consistently with the Area-Entropy Law.

\section{\normalsize  Concluding Remarks  and Further Outlook}

 In the present paper, we first emphasized the importance of  the noncommutative space-time towards ultimate theory of quantum gravity and Planck scale physics.  Indeed, {\it entirely} on the basis of Kinematical Holographic Relation [KHR] intrinsic in the Yang's quantized space-time, we have successfully come close to the fundamental problems of black hole entropy, that is, our new area-entropy relation in section 4 and the unification of The first law of black hole mechanics and The thermodynamics of black holes in the preceding section. 

It should be noted here that first our new area-entropy relation is constructed {\it statistically} in the familiar way inside $V_3^{R_S}$ through  [KHR],  and second the holographic structure of our area-entropy relation clearly stems from the holographic structure of  [KHR] itself, $ n_{\rm dof}  (V_d^L)= {\cal A} (V_d^L) / G_d$. \footnote {Furthermore, one should notice that the kinematical holographic relation [KHR],   $ n_{\rm dof}  (V_d^L)= {\cal A} (V_d^L) / G_d,$ has a possibility of providing a new substantial basis for the familiar arguments around the so-called AdS/CFT correspondence problem [23] or the  Entanglement entropy hypothesis of black hole entropy, both related to the bulk and boundary of $V_d^L,$ without difficulties of infrared- and ultraviolet divergences.[38] } 

In addition, let us notice the special structure of our area-entropy relation [AER] $S_S (V_3^L) = {\cal A} (V_3^{R_S}) S_S[site] / 4\pi $ (4.20), together with the relation  $\eta = S_S[site] / 4 \pi$ (4.21) obtained in section 4, although our scheme could not determine the exact value of  Bekenstein parameter $\eta.$ \footnote {It is well known that $\eta$ was successfully calculated by Strominger and Vafa  [15] with the specific value   $\eta =1/4$, in precise agreement  with the familiar Bekenstein-Hawking's area-entropy relation. Indeed, it gives us an important clue to make clear Planck scale physics, even though it also has its own problems to be solved.[16] }  
It is quite important, however, to notice that  $\eta$ is expressed in terms of $S_S[site]$, that is, $\eta = S_S[site] / 4 \pi$, so we have now a possibility of clarifying   its physical meaning  in the framework of our present scheme. 

Let us first focus our attention on $S_S[site]$. As a matter of fact, we have referred to it in the context of a certain kind of {\it universality} of black holes in subsection 4.2. At this point, it is important to notice the relation  
\begin{eqnarray}
S_S(V_3^{R_S}) = n_{\rm dof}(V_3^{R_S}) S_S[site],   
\end{eqnarray}
which is simply derived from (4.10) under $L= R_S (=2M(V_3^{R_S})).$  The relation suggests that $S_S[site]$ represents a kind of {\it universal} unit of entropy of black holes, which appears as the entropy realized on each individual [site] inside any black hole, by taking a  proper specific value, $S_S[site] = 4\pi \eta$.

Next, let us turn our attention to ${\bar N}_S[site] $.  Indeed, it has also a certain universal nature as $S_S[site]$ on account of the relation 
\begin{eqnarray}
S_S [site]\  = \ln (1+ {\bar N}_S[site]) + {\bar N}_S [site] \ln (1+ {\bar N}_S^{-1}[site],
\end{eqnarray}
which is derived from (4.17) in section 4. One immediately finds out that  ${\bar N}_S[site] $ must be also a function of $\eta$ in accord with $S_S[site] =4\pi \eta$ and denotes the average number of $D_0$ brane inside each individual [site] in association with $S_S[site]$ mentioned above. Indeed, it gets a fixed value under a pecific value of $\eta.$ For instance,  when  $\eta =1/4$,  
\begin{eqnarray}
{\bar N}_S[site] \sim 1/0.12.
\end{eqnarray} 

Now, under the above two {\it universal} quantities, $S_S[site]$ and ${\bar N}_S[site]$, let us consider black holes in general with any total mass $M_S,$ noticing the relation
\begin{eqnarray}
\mu_S  {\bar N}_S[site]  =1/(4 M_S  ),
\end{eqnarray}  
which is simply derived by rewriting the expression  [BHC] (4.19). It shows  that the effective mass at each [site], $\mu_S  {\bar N}_S[site], $ decreases inversely  with the total mass $M_S$ as $1/(4 M_S )$. On the other hand, the same relation (6.4) under (6.3) gives   
\begin{eqnarray}
\mu_S (= 1/(4{\bar N}_S[site] M_S)) \sim 0.03/M_S.
\end{eqnarray}
Taking into consideration that  $\mu _S$ denotes by definition the effective mass of $D_0$ brane inside a black hole with total mass $M_S,$  one finds that the relation (6.5) can be significantly compared with (4.26) concerning  $T_{H.R.}$ under  the same condition $\eta =1/4,$
\begin{eqnarray}
T_{H.R.} (= 1/(8 M_S S_S[site]) ) \sim 0.04/M_S.  
\end{eqnarray}

These results derived of the {\it universal} nature of black holes seem suggestive for Planck scale physics.

In conclusion, we emphasize again the significance of Yang's quantized space-time, which first of all gives us a possibility of being free from the well-known difficulty of singularities [8-9],[38],  as was pointed out in Introduction. On the other hand, in the present paper, we finally arrived at an important result of {\it statistical} and {\it substantial} understanding of area-entropy law of black holes under a novel concept of noncommutative quantized space-time, through the kinematicval holographic relation [KHR] intrinsic in Yang's quantized space-time.  

We anticipate that noncommutative quantized space-time will make breakthrough towards the ultimate theory of quantum gravity and Planck scale physics,  as  Planck's Quantum Hypothesis (1900) once played the ultimate role in clearing away ``Nineteenth century clouds over the dynamical theory of heat and light" towards quantum physics.       

As was remarked in Introduction,  kinematical reduction of spatial degrees of freedom, which underlies [KHR], is considered to hold widely in the noncommutative space-time, so it is important to examine how the kinematical reduction of spatial degrees of freedom may occur in the noncommutative space-time in general.  Furthermore, it is our important task to improve the simple $D_0$ brane gas model in the present paper,  towards a reconstruction of M-theory [20] in the noncommutative quantized space-time [21]-[22], [35]-[38] or Planck scale physics.$^7$

\vskip 1cm
{\normalsize \bf  Acknowledgment}
\vskip 0.5cm
I would like to thank  H.~Aoyama for giving constant encouragement to my work and for  careful reading of  the present manuscript and valuable comments.

\end{document}